\documentclass[journal=achre4,manuscript=review,layout=twocolumn]{achemso}
\usepackage{graphicx}
\usepackage{amsmath,amssymb,amsfonts}
\usepackage{bm}
\usepackage[version=3]{mhchem}
\usepackage[bookmarks=false]{hyperref}
\usepackage{xcolor}
\usepackage{xfrac}

\hypersetup{
  pdfstartview = FitH,
  pdftoolbar   = false,
  pdfmenubar   = true,
  colorlinks   = true,
  linkcolor    = blue!50!black,
  urlcolor     = blue!50!black,
  citecolor    = blue!50!black
}

\title{Chemistry in Motion: Tiny Synthetic Motors}
\author{Peter H. Colberg}
\author{Shang Yik Reigh}
\author{Bryan Robertson}
\author{Raymond Kapral}
\email{rkapral@chem.utoronto.ca}
\affiliation{Chemical Physics Theory Group, Department of Chemistry,\\
University of Toronto, Toronto, Ontario M5S 3H6 Canada}

\begin{document}

\begin{abstract}
Diffusion is the principal transport mechanism that controls the motion
of solute molecules and other species in solution; however, the random
walk process that underlies diffusion is slow and often nonspecific.
Although diffusion is an essential mechanism for transport in the biological
realm, biological systems have devised more efficient transport mechanisms
using molecular motors.
Most biological motors utilize some form of chemical energy derived from
their surroundings to induce conformational changes in order to carry
out specific functions.
These small molecular motors operate in the presence of strong thermal
fluctuations and in the regime of low Reynolds numbers, where viscous
forces dominate inertial forces.
Thus, their dynamical behavior is fundamentally different from that of
macroscopic motors, and different mechanisms are responsible for the
production of useful mechanical motion.

There is no reason why our interest should be confined to the small motors
that occur naturally in biological systems.
Recently, micron and nanoscale motors that use chemical energy to produce
directed motion by a number of different mechanisms have been made in
the laboratory.
These small synthetic motors also experience strong thermal fluctuations
and operate in regimes where viscous forces dominate.
Potentially, these motors could be directed to perform
different transport tasks, analogous to those of biological motors,
for both {\it in vivo} and {\it in vitro} applications.
Although some synthetic motors execute conformational changes to effect
motion, the majority do not, and, instead, they use other mechanisms to
convert chemical energy into directed motion.

\begin{figure}[tb]
  \centering
  \includegraphics[width=\linewidth]{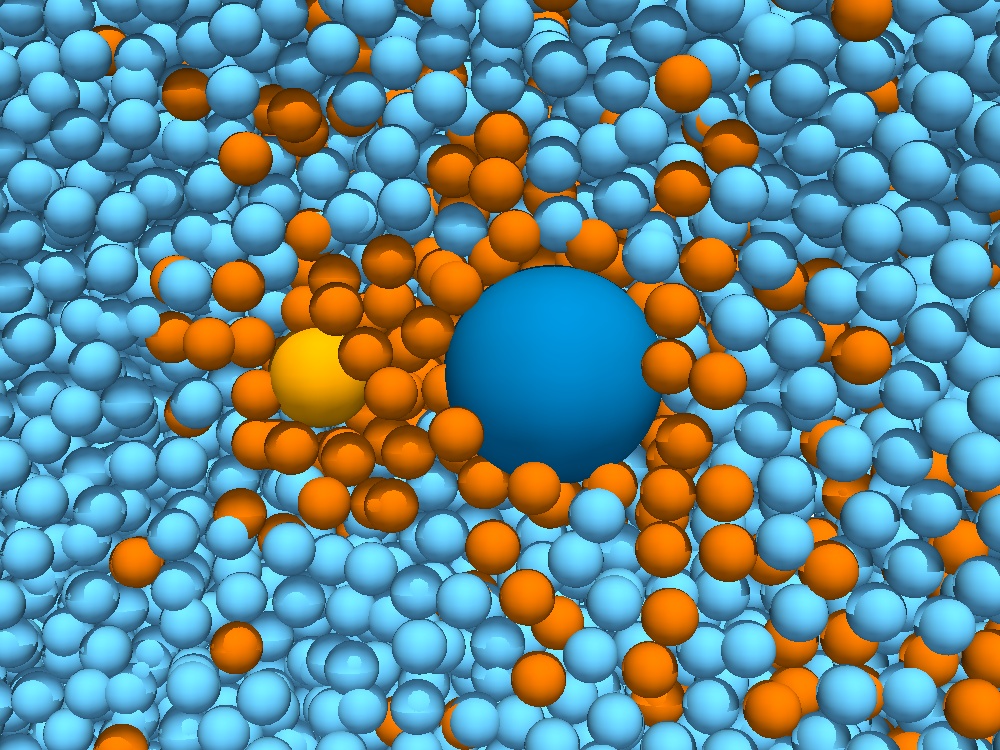}
\end{figure}
In this Account, we describe how synthetic motors that operate by
self-diffusiophoresis make use of a self-generated concentration gradient
to drive motor motion.
A description of propulsion by self-diffusiophoresis is presented
for Janus particle motors comprising catalytic and noncatalytic faces.
The properties of the dynamics of chemically powered motors are
illustrated by presenting the results of particle-based simulations of
sphere-dimer motors constructed from linked catalytic and noncatalytic
spheres.
The geometries of both Janus and sphere-dimer motors with asymmetric
catalytic activity support the formation of concentration gradients
around the motors.
Because directed motion can occur only when the system is not in
equilibrium, the nature of the environment and the role it plays in
motor dynamics are described.
Rotational Brownian motion also acts to limit directed motion, and it has
especially strong effects for very small motors.
We address the following question: how small can motors be and still exhibit
effects due to propulsion, even if only to enhance diffusion?
Synthetic motors have the potential to transform the manner in which
chemical dynamical processes are carried out for a wide range of
applications.
\end{abstract}

\maketitle

\section{Introduction} \label{sec:intro}
A large effort has been made in recent years to synthesize and study
micron and nanoscale motors and machines.
These small devices are constructed from a wide range of materials, such
as DNA, polymers, or metals, take many different shapes and sizes, and,
owing to such variety, are able to move using different mechanisms.
They have been shown to act, at least in proof-of-concept, as walkers,
shuttles, rotors, pumps, cargo carriers, muscles, and artificial flagella
and cilia.~\cite{kay:07,jones-book,wangbook:13}
Some synthetic motors are designed to move as a result of nonreciprocal
cyclic conformational changes, mimicking many of the motors and
microorganisms found in nature.
Motors that do not rely on conformational changes for motion have also
been constructed, and these motors are the principal focus of this Account.

Apart from nanomotors without moving parts that are propelled by
external stimuli, many such motors rely on chemical reactions for
propulsion.~\cite{kapral:13}
The fuel they use is not carried by the motor itself, but, rather, it
is derived from the local environment: they are active devices that use
the chemical energy available in their immediate vicinity to perform work.
In place of asymmetrical conformation changes, these chemically powered
motors are constructed with a structural asymmetry in chemical activity
that leads to directed motion.
The motor motion depends on the characteristics of both the motor and
its environment and their interaction with one another.

The first chemically propelled nanomotors were bimetallic rods constructed
with gold and platinum or nickel portions, which, when placed in solutions
containing hydrogen peroxide, were able to execute autonomous linear and
rotational motions.~\cite{sen:04,ozin:05}
Subsequently, other types of chemically driven motors were
synthesized.~\cite{wangbook:13,wang:13}
Similar to bimetallic rods, these motors contain two main domains that
are composed of different materials, allowing for chemical reactions to
occur asymmetrically on the motor.
Janus particles are spherical colloids generally made from one
material, half covered by another material, giving the appearance of
two faces.~\cite{golestanian-1:07}
Sphere-dimer motors are similar to Janus particles, but they comprise
two linked spheres made of different materials.~\cite{ruckner:07,ozin:10}
Microtubular motors are hollow cylindrical or conical tubes that have
a reactive interior and inert exterior.~\cite{mei:08,soler:13,zhao:14}
Although many of these motors rely on the same platinum--peroxide catalytic
reaction, motors constructed more recently have incorporated different
metals, alloys, or compounds in order to provide better control over
motion or to exploit different fuels.~\cite{wangbook:13}

Regardless of the specific motor geometry or fuel, two classes of
mechanisms are largely responsible for the self-propulsion of
chemically driven nanomotors.~\cite{wangbook:13,kapral:13}
Bubble propulsion is due to the catalytic production of gas at the motor
and the resulting recoil of the gas bubbles from the motor surface.
This mechanism is usually invoked whenever bubbles are directly observed,
as is the case with tubular motors, where gas accumulates in
the motor interior and is expelled from one of the ends.
In phoretic mechanisms, the motor self-generates some type of gradient
(electrical, concentration, temperature) in its vicinity through a
chemical reaction, and motion is induced by this gradient.
For bimetallic rod motors, a flow of electrons in the rod between the
anodic and cathodic sites of a redox reaction creates a self-generated
electric field that drives ion motion in the electrical double layer
surrounding the motor.~\cite{wang:06}
In some situations, more than one phoretic mechanism may operate for the
same motor.
For example, the mechanism by which sphere dimers move is highly dependent
on the rate of catalysis and surface roughness and may move either by
bubble propulsion or self-diffusiophoresis.~\cite{wu:14}
Both self-electrophoretic and self-diffusiophoretic mechanisms
have been shown to contribute to the propulsion of the same
motor.~\cite{ebbens:14,brown:14}
Simulations of Janus particles fueled by exothermic reactions have also
shown that self-diffusiophoresis and self-thermophoresis may act at the
same time and possibly in opposite directions.~\cite{debuyl:13}

\section{Chemically Powered Motors} \label{sec:mechanism}
\begin{figure}[tb]
  \centering
  \includegraphics[scale=0.6]{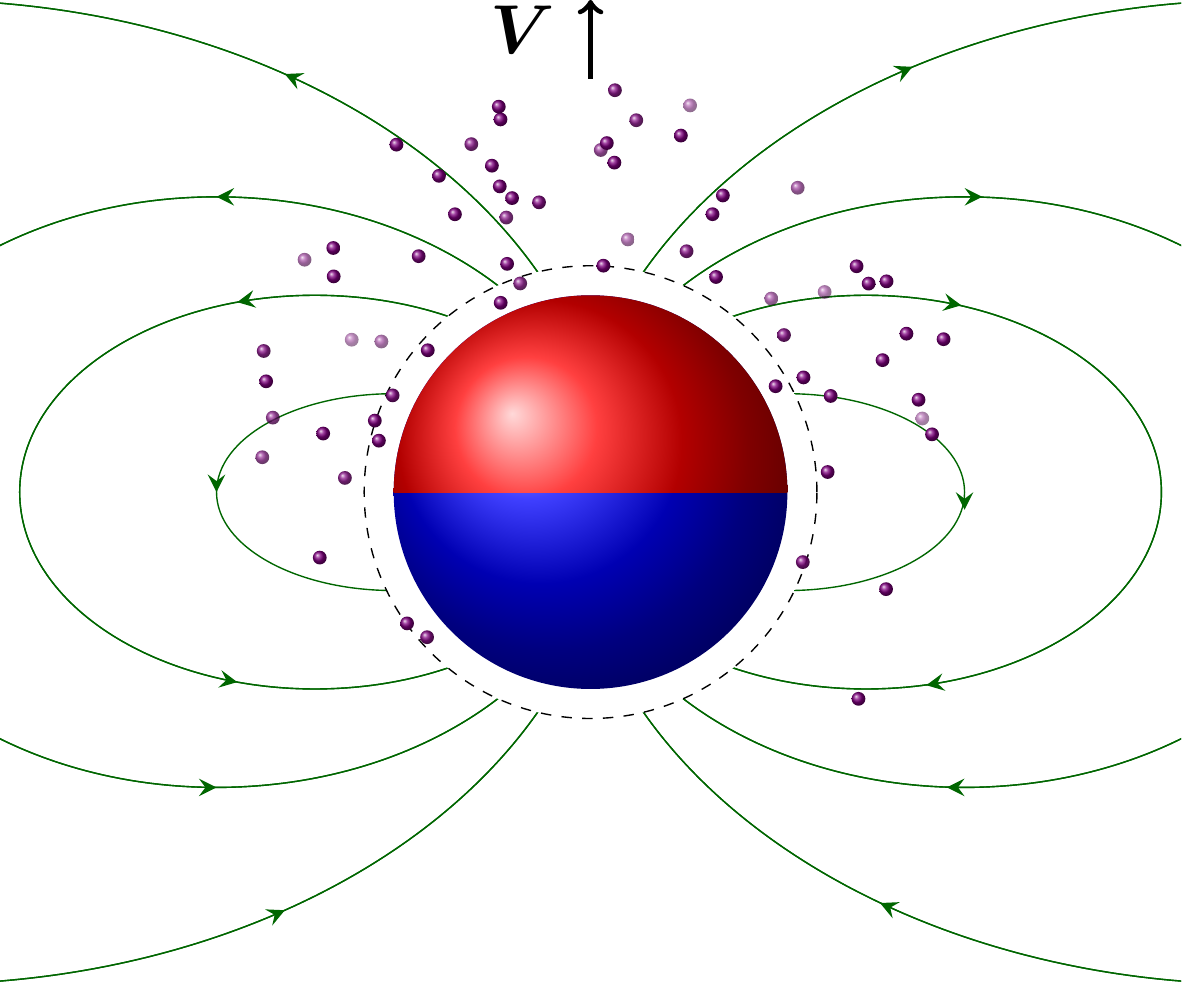}
  \caption{A Janus particle, in the laboratory frame of reference,
  which is self-propelled by a diffusiophoretic mechanism.
  The catalytic part (red, upper hemisphere) catalyzes the conversion
  of the fuel A molecules (not shown) to the product B molecules
  (small purple spheres).
  The chemically inactive face is the lower (blue) hemisphere.
  The system parameters are chosen so that $\bm{V}$ points from the
  N to the C face.
  The dipolar fluid velocity field in the particle vicinity is also shown
  (green lines).}
  \label{fig:janus}
\end{figure}

The means by which small motors use chemical reactions to execute directed
motion may be illustrated by considering a spherical Janus particle with
catalytic (C) and noncatalytic (N) faces, which is shown in
Figure~\ref{fig:janus}.
The dynamics of the Janus particle depends on the specific chemical
reactions that are responsible for self-propulsion.
Here, we suppose that a simple idealized \ce{A -> B} reaction takes place on the
C face of the Janus particle.
This model will still account for many of the generic features of real motor
motion.
The fluid surrounding the particle contains solvent S, fuel A,
and product B chemical species.

The molecular origin of propulsion can be traced to the different
interactions that the fuel and product molecules have with the Janus
particle.
The intermolecular potentials are assumed to be short-ranged and take
non-zero values only within a thin interfacial region surrounding the
Janus particle (dashed line in Figure~\ref{fig:janus}).
Because the catalytic activity is confined to one face of the particle,
a nonuniform distribution of fuel and product molecules is produced in
its vicinity.
The resulting concentration gradient gives rise to a force on the Janus
particle.
Because there are no external forces acting on the system and the
intermolecular forces have short range, the Janus particle plus the
solvent within the boundary layer is force-free.
Through momentum conservation, a flow is generated in the surrounding fluid,
and the particle is propelled in a direction opposite to the fluid flow.
From these qualitative considerations, detailed theoretical
expressions for the motor velocity, based on a continuum description of the
motor environment, have been given previously in the
literature.%
~\cite{golestanian:05,golestanian:07,julicher:09,popescu:10,sabass:11,sharifi-mood:13,lauga:14}
Below, we present an outline of such theoretical results, specialized to the
simple \ce{A -> B} reaction model.

The fluid flow in the boundary region leads
to a slip velocity, $\bm{v}_\text{s}$, on its outer edge at $r=R_0$, whose
value depends on the concentration gradient, intermolecular forces, and
solvent viscosity through the relation~\cite{anderson:86,anderson:89}
\begin{equation}
  \bm{v}_\text{s} (R_0,\theta)=
  \frac{k_\text{B}T}{\eta}\left(\Lambda_\text{N}
  +(\Lambda_\text{C}-\Lambda_\text{N}) H(\theta)\right)
  \nabla_\theta c_\text{A}(R_0)
  \label{slip}
\end{equation}
where $\theta$ is the polar angle in a spherical polar coordinate system,
$k_\text{B}$ is the Boltzmann constant, $\eta$ is the shear viscosity, $T$
is the absolute temperature, $c_\text{A}$ is the concentration of the A
molecules, and the function $H(\theta)$ is defined to be unity on the
catalytic C hemisphere ($0\leq \theta \leq\sfrac{\pi}{2}$) and zero on the
noncatalytic N hemisphere ($\sfrac{\pi}{2}<\theta \leq \pi$).%
~\cite{golestanian:07}
In this equation, the intermolecular potentials enter through $\Lambda_I$
\begin{equation}
  \Lambda_\text{I} = \int_0^\infty dr \; r\left(e^{-\beta U_\text{BI}(r)}
  -e^{-\beta U_\text{AI}(r)}\right)
\end{equation}
where $\beta=1/(k_\text{B}T)$ and $U_{\alpha\text{I}}$ is the potential of mean
force between the chemical species $\alpha$ ($\alpha=\text{A,B}$) and the I
($I=\text{C,N}$) hemisphere of the Janus particle.
The velocity of the Janus particle is given in terms of the surface
(${\mathcal S}$) average of the slip velocity as
$\bm{V}=-\langle\bm{v}_\text{s}\rangle_{{\mathcal S}}$, where
$\langle\bm{v}_\text{s}\rangle_{{\mathcal S}}=\int_{{\mathcal S}} \bm{v}_\text{s} d{\mathcal S}/(4\pi R_0^2)$.
The $\Lambda_\text{I}$ parameters can take either sign, and this determines the
direction of propagation.

The concentration field of the A molecules is found by solving
the steady-state diffusion equation, $ \nabla^2 c_\text{A}=0$, subject to
a reflecting boundary condition on the noncatalytic portion of the
surface at $R$ ($\approx$$R_0$) and a radiation boundary condition
on the catalytic surface: $(D\hat{\bm{r}} \cdot \nabla c_\text{A} )_{r=R}
=\bar{k}_0 c_\text{A}(R)H(\theta)$, where $D$ is the relative diffusion constant
of A and the Janus particle, $\hat{\bm{r}}$ is the unit vector along
$\bm{r}$, $\bar{k}_0=k_0/(4\pi R^2)$, and $k_0$ is the intrinsic reaction
rate constant.
Far from the particle, we assume there exists only fuel with concentration
$c_0$, so $c_\text{A}(r \to \infty)=c_0$.
The Janus particle velocity is found by substituting the concentration
field obtained from the solution of the diffusion equation into the
slip-velocity equation and then taking the surface average as indicated
above.
The result for $V_z=\hat{\bm{z}}\cdot \bm{V}$, the motor velocity
along the unit vector, $\hat{\bm{z}}$, from the center of the Janus
particle to the pole of the C hemisphere is
\begin{align}
  V_z = \frac{k_\text{B}Tc_0 \bar{k}_0}{\eta D}(a_\text{C} \Lambda_\text{C}
  +a_\text{N} \Lambda_\text{N}),
\end{align}
where $a_\text{I}$ ($\text{I}=\text{C,N}$) are coefficients determined from the
solution of the reaction-diffusion equation and depend on the ratio $k_0/k_D$,
with $k_D=4\pi R D$ the Smoluchowski rate coefficient.

This expression can be used to determine how $V_z$ varies with system
parameters.
For instance, consider changing the radius $R$ of the Janus particle.
The $\Lambda_I$ terms depend on the boundary layer thickness $\delta$ and
particle size and vary linearly with $R$ for $R \gg \delta$.
Also, $k_0 \sim R^2$, $\bar{k}_0$ is independent of $R$, and $k_D \sim R$.
The full rate coefficient for the \ce{A -> B} reaction may be written as
$k = k_0 k_D/(k_0 + k_D)$.
Then, for small Janus particles, $k_0/k_D \ll 1$, $k \approx k_0$, and we
have reaction controlled kinetics.
In this limit, one may show that the $a_\text{I}$ coefficients are independent
of $k_0/k_D$ and thus the $R$ dependence of $V_z$ is controlled by the
$\Lambda_I$ terms.
If, instead, the particle is large, then $k_0/k_D \gg 1$, $k \approx k_D$,
and we have diffusion-controlled kinetics.
The $a_\text{I}$ coefficients now depend on $R$, and the nature of the $R$
dependence requires the specific form of the solution of the
reaction-diffusion equation.
The precise value of $R$ where the kinetics changes from reaction to
diffusion control depends on the intrinsic reaction rate and species
diffusion coefficients.
These comments, along with earlier work on the size dependence of Janus
particle velocity~\cite{golestanian:12}, indicate that the motor velocity
may vary in nontrivial ways as the system parameters are changed.

The fluid flow outside the interfacial layer can be calculated from the solution
to the Stokes equation, and in the laboratory frame of reference, it is given by
$\bm{v}(\bm{r})=\tfrac{1}{2}(R/r)^3(3\hat{\bm{r}}\hat{\bm{r}}-\bm{I})\cdot\bm{V}$.%
~\cite{anderson:86,anderson:89}
The fluid-flow lines around the self-propelled Janus particle are shown
in Figure~\ref{fig:janus}.
The dipolar form of these flow lines implies a $1/r^3$ decay of the
velocity field far from the particle.

\begin{figure}[tb]
  \centering
  \includegraphics[scale=0.6]{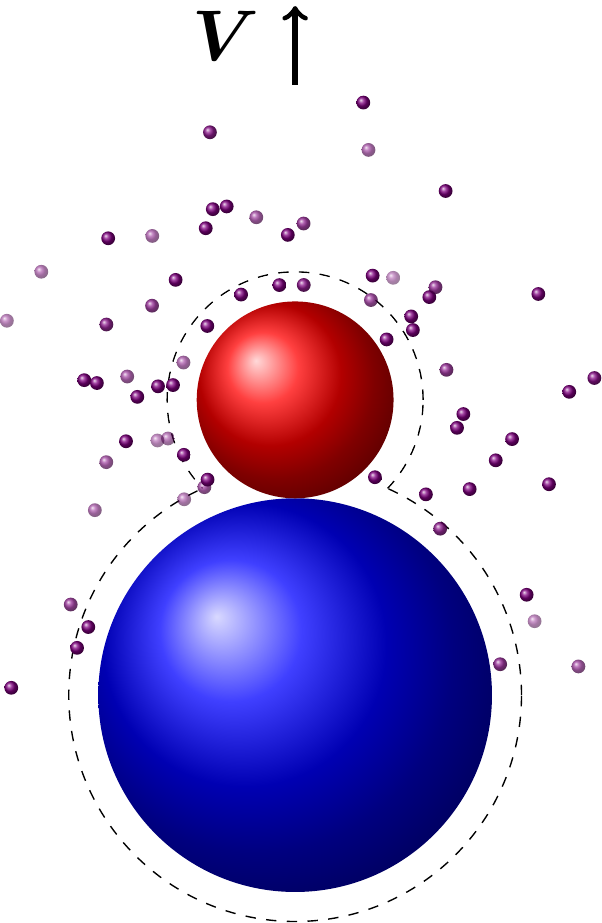}
  \caption{A sphere-dimer motor propelled by self-diffusiophoresis: the
  catalytic sphere (red) converts fuel molecules (not shown) to product
  molecules (small purple spheres).
  The noncatalytic sphere is depicted in blue, and the interfacial region
  is shown as a dashed line.
  The system parameters are again chosen so that $\bm{V}$ points from
  the N to the C sphere.}
  \label{fig:dimer}
\end{figure}

Another simple motor geometry consists of two linked catalytic and
noncatalytic spheres (see Figure~\ref{fig:dimer}).
An analysis similar to that described above for Janus particles, but
technically more involved, can be carried out for these motors to obtain
the velocity~\cite{popescu:11}.

Similar analyses can be carried out for propulsion by other self-phoretic
mechanisms, although some details of the calculation differ.
It is important to recognize that our Janus particle example was idealized
and, in applications to motor motion in the laboratory, details of the
reaction mechanism and other factors may need to be taken into
account.~\cite{sabass:11,golestanian:12}
Bimetallic rod and Janus particle motors operate by self-electrophoresis,
and the precise nature of the oxidation and reduction reactions that
take place on the motor will influence its dynamics.~\cite{wang:06}

\section{Nanomotor Dynamics}\label{sec:complex}
Small chemically propelled motors are strongly influenced by thermal fluctuations.
Also, as the motor size decreases to nanometer scales, the validity
of macroscopic models for the dynamics should be examined to determine
their applicability.
For these reasons, it is appropriate to consider particle-based descriptions
where the dynamics of the entire system is described by either molecular
dynamics or mesoscopic dynamical schemes that retain the important features of
full molecular dynamics.
It is especially important to preserve momentum conservation for the
reasons described earlier.
The results that follow were derived from simulations using hybrid
molecular dynamics (MD)-multiparticle collision dynamics
(MPCD)~\cite{malevanets:99,malevanets:00,kapral:08,gompper:09} or full MD,
depending on the size of the motor being studied.
The full MD simulations were carried out for a simple model system but
could be extended to treat specific real systems.
By contrast, the larger-scale MD-MPC simulations utilize coarse-grained
descriptions of all species and additionally neglect solvent structural
effects.
Nevertheless, these simulation models account for the principal elements of
self-diffusiophoretic motion.

Self-propulsion can occur only under nonequilibrium conditions since
detailed balance prohibits directed motion in an equilibrium system.
For sustained motion to occur, the system must be maintained in a
nonequilibrium state by fluxes of reagents at the boundaries or by bulk
reactions that are themselves forced out of equilibrium.

The examples of chemically powered motor dynamics given below will be
confined to sphere-dimer motors since most of our simulations have been
performed with this motor geometry~\cite{yuguo:08}; however,
many of the phenomena we describe are observed for other motor geometries.
The investigation of the dynamics of synthetic self-propelled nanomotors is
an active area of research, and there is a large amount of literature describing
work in this area.~\cite{jones-book,wangbook:13}
A survey of this literature is beyond the scope of this Account.
Instead, we shall highlight just a few aspects of motor dynamics that are
dictated by our current interests: motor motion in complex media,
collective motor motion, and the dynamics of very small molecular-scale
motors.

In the biological realm, molecular motors operate in complex
nonequilibrium environments where the surrounding medium supports networks
of chemical reactions that supply fuel and remove product.
Synthetic, chemically-powered motors may also operate in such complex
chemical media, and we present two examples to illustrate the new
phenomena that arise in such situations.

\begin{figure}[tb]
  \centering
  \includegraphics[width=\linewidth]{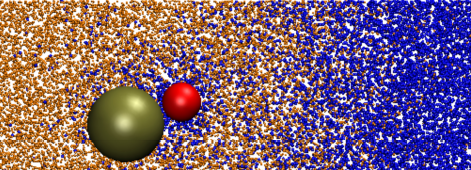}
  \caption{Propagating chemical wave and sphere dimer at one time instant.%
  ~\cite{snigdha:11}
  Species A and B are rendered in red and blue, respectively.
  The chemical front and the self-generated concentration field around
  the motor can bee seen in this figure.}
  \label{fig:wave}
\end{figure}

Consider a sphere-dimer motor where the reaction \ce{A -> B} on the
catalytic sphere generates the chemical gradient responsible for
propulsion.
The medium in which the motor moves supports the nonequilibrium cubic
autocatalytic reaction, \ce{B + 2A -> 3A}, where A is the autocatalyst.%
~\cite{snigdha:11}
Notice that the bulk reaction consumes the product and regenerates the
fuel so that motor motion may be sustained.
This bulk reaction also supports the formation of a traveling chemical
wave: if half of the system is initially filled with A species and the
other half with B, then the autocatalyst will consume the B particles at
the interface between these two regions, leading to the formation of a
propagating front.
A sphere-dimer motor placed in the fuel-rich A domain and oriented
toward the interface will encounter the front provided its speed is
greater than that of the front.
Because the system is rich in product B beyond the front, the motor
cannot penetrate deeply into this fuel-poor region.
If the motor were oriented perpendicular to the front and
orientational Brownian motion were suppressed, then the motor would propagate
with the front; however, Brownian reorientation and self-propulsion will
cause the motor to re-enter the fuel-rich A domain, giving rise to
reflection-like dynamics in the front vicinity (see Figure~\ref{fig:wave}).
This feature suggests the possible control of motor motion by chemical
patterns.

\begin{figure}[tb]
  \centering
  \includegraphics[width=\linewidth]{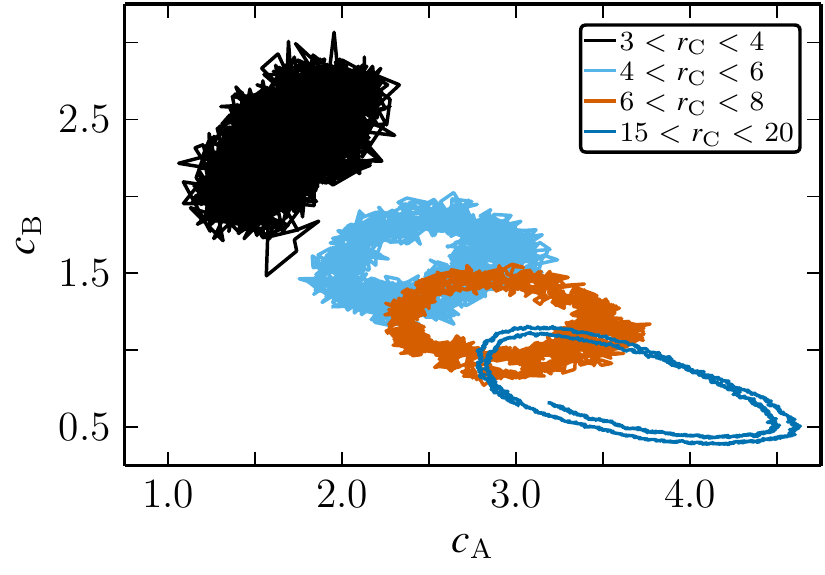}
  \caption{Concentrations of A and B are plotted over the course
  of several cycles.
  The different concentration plots correspond to the concentration
  in spherical shells around the catalytic sphere at different radii,
  $r_\text{C}$.}
  \label{fig:cycles}
\end{figure}

The medium in which the motor moves can support even more complex
nonequilibrium oscillatory states.
Oscillatory dynamics is commonly observed in biological systems where
coupled autocatalytic reactions give rise to the periodic behavior.
To study motor dynamics in such media, we again suppose that the reaction
on the catalytic sphere of the sphere-dimer motor is \ce{A -> B}.
These species are also involved in bulk nonequilibrium reactions whose
kinetics is controlled by the Selkov model: \ce{S <=> A}, \ce{A + 2B <=> 3B},
\ce{B <=> S}, where S is considered to be an inert feed for A and B.
The rate constants in these reactions can be chosen to yield an
oscillatory state.
The Selkov model has its antecedents as a simple model for glycolytic
oscillations.
Since the reaction on the dimer motor involves the same chemical species,
it locally perturbs the Selkov oscillatory dynamics.
In particular, the concentration of the product B species is observed
to oscillate around a higher average value close to the
catalytic sphere, whereas the opposite trend is seen for the fuel A
concentration (see Figure~\ref{fig:cycles}).
These shifts in the concentration cycles also create oscillations in
the concentration gradient across the noncatalytic sphere, which lead
to oscillations in the sphere dimer's velocity.
Thus, the motor is able to influence the local chemical kinetics of an
oscillatory medium and, in turn, these changes modify the motor motion.

Next, we briefly consider the collective motion of chemically powered
motors.
Systems containing many motors constitute active media, which have been
shown to exhibit phenomena that are quite distinct from those seen in
systems with inactive components.
Active systems exist in out-of-equilibrium states, and the investigation of
the properties of this new class of systems is a rapidly growing research
area. ~\cite{sriram-review:13}
The active constituents can have diverse length scales, ranging from the
large macroscopic sizes of birds in flocks or fish in schools, to
micron-sized swimming organisms.
In these examples, the active objects are powered internally but are
sustained by the input of food sources.
Models for such systems often assign a velocity to individual active
objects and interactions of various kinds among the objects lead to
nontrivial collective behavior.

The collective motion of chemically powered motors has been studied
experimentally~\cite{bocquet:12,speck:13} and
theoretically~\cite{snigdha:12,soto:14,golestanian:14}.
Phenomena, such as active self-assembly into dynamical clusters with various
geometries and swarming behavior, have been observed.
Several factors have to be taken into account when the dynamics of an
ensemble of chemically powered motors is considered.
Each self-propelled motor generates its own concentration gradient.
This gives rise to a chemotactic response that typically causes motors to
be attracted to each other, analogous in some respects to the chemotactic
response of bacteria to high food concentrations~\cite{berg}.
The chemotactic response of synthetic nanomotors to chemical gradients
has been observed in experiments.~\cite{hong:07,baraban:13}
Also, each motor induces a flow field in the surroundings, and one motor can
perturb the flow field of its neighbors, leading to a hydrodynamic coupling
among motors.
The motors may also interact directly through short- or long-range
interactions, and because of all of these interactions, the collective
dynamics may be complicated.

\begin{figure}[tb]
  \centering
  \includegraphics[width=\linewidth]{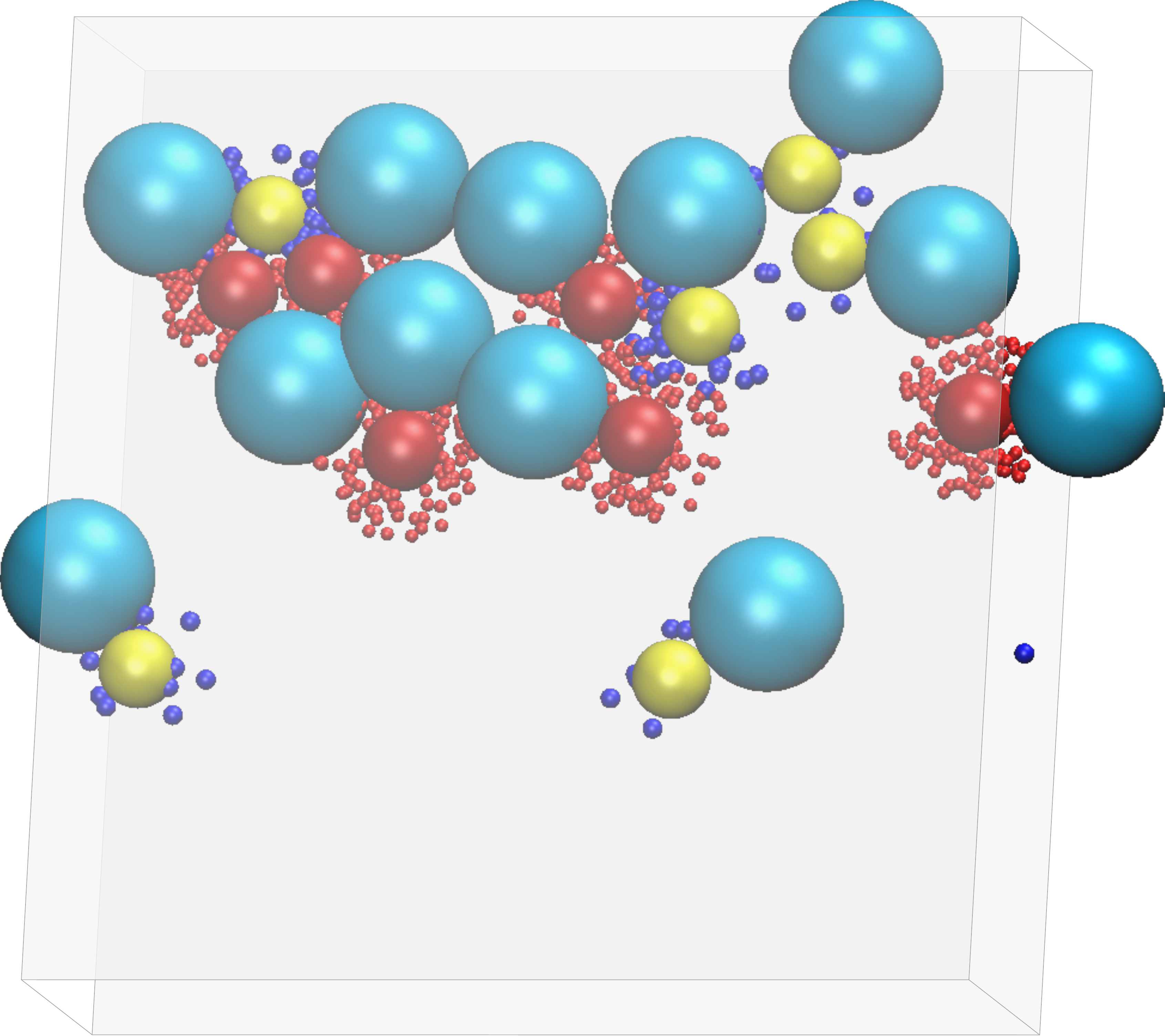}
  \caption{An instantaneous configuration showing the collective behavior
  of a mixture of types I (red catalytic sphere) and II (yellow catalytic
  sphere) sphere-dimer nanomotors.
  The product species B (red) and C (blue) are shown only in the
  immediate vicinities of the motors.}
  \label{fig:many}
\end{figure}

Earlier studies of the collective behavior of chemically powered
motors considered ensembles of identical motors.%
~\cite{snigdha:12,soto:14,golestanian:14}
An even richer phenomenology is found when the ensemble consists of motors
of two types that can couple through chemical reactions.
Consider an ensemble of types I and II sphere-dimer motors, which
interact through chemical gradients in the following way: the product
of a type I motor is the fuel for a type II motor.
In particular, type I motors catalyze the reaction \ce{A -> B}, whereas
type II motors catalyze the reaction \ce{B -> C}.
This is an example of a system where the motors themselves participate
in networks of chemical reactions.
If the system is supplied with fuel A, then motors of type II will not
actively move unless they are in the vicinity of motors of type I, which
provide their fuel.
An instantaneous configuration that is formed in the course of the
evolution is shown in Figure~\ref{fig:many}.
The type I motors actively aggregate into dynamical clusters, and the type
II motors tend to aggregate on the surfaces of these clusters. 
Both theoretical and experimental investigations of the collective
behavior of chemically powered motors are at an early stage, and significant
progress is anticipated in this area.

\section{Fluctuations and Diffusion} \label{sec:next}

\begin{figure}[tb]
  \centering
  \includegraphics[width=\linewidth]{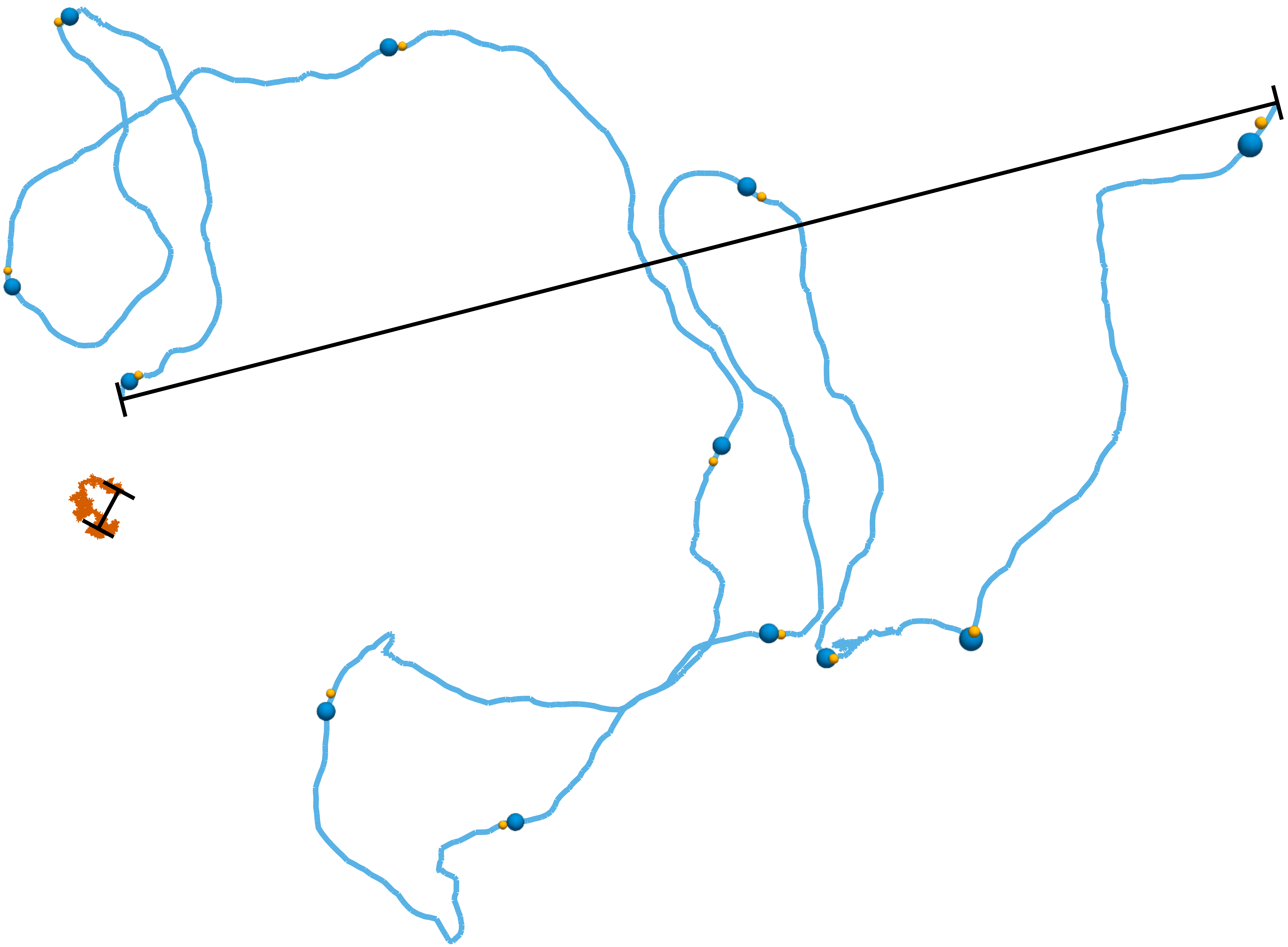}
  \caption{Trajectory of a nanoscale sphere-dimer motor (sky blue).
  For comparison, a trajectory of the same duration is shown for an
  inactive dimer (no motor reactions) is also shown (vermilion).
  The reactive dimer travels a 40~times greater end-to-end distance
  (black) than the inactive dimer.}
  \label{fig:dimer-traj}
\end{figure}

While the chemical concentration gradient
generated by the motor determines its propagation direction and affects
its speed, orientational Brownian motion will change the direction in
which it moves so that on sufficiently long time scales the directed
motion will manifest itself as enhanced diffusion.
The enhanced diffusion is evident from a comparison of the two
trajectories in Figure~\ref{fig:dimer-traj}, corresponding to chemically
active and inactive dimers.
Only for times less than the orientational time will the ballistic motion
of the motor be evident.
A number of important questions arise when the effects of fluctuations
on motor dynamics are considered.
How far, on average, does the motor move before the ballistic motion is
masked by Brownian motion? How is the diffusion coefficient modified
by directed motion? What is the lower limit on the size of the motor
for self-diffusiophoresis to operate? For micron and large nanoscale
motors, the answers to these questions will determine how effectively the
motor can carry out transport tasks and what control scenarios must be
implemented to overcome the effects of rotational Brownian motion.
If such motors are ever to be used on scales comparable to the interior
of a cell, it is important to determine if they can be made to operate
when they are only a few nanometers in size.

The mean square displacement (MSD), $\Delta L^2(t)$, of the center of
mass of the motor provides some information that can be used to answer
these questions.
We may write the motor velocity as the sum of the average velocity
along the instantaneous bond unit vector $\hat{{\bm z}}(t)$, and a
fluctuation $\delta\bm{V}(t)$, $\bm{V}(t) =
\langle V_z\rangle \hat{{\bm z}}(t) + \delta\bm{V}(t)$.
Assuming exponential decay for the orientational
$\langle \hat{{\bm z}}(t) \cdot \hat{{\bm z}}
\rangle = {\mathrm e}^{-t/\tau_\text{r}}$ and velocity fluctuation
$\langle\delta\bm{V}(t)\cdot\delta\bm{V}\rangle =
(3k_\text{B}T/M_\text{m})\,{\mathrm e}^{-t/\tau_\text{v}}$ correlation functions,
the MSD takes the form
\begin{eqnarray}
  \label{eq:MSD-th}
  &&\Delta L^2(t)
   = 6D_\text{m}t
  - 2\langle V_z\rangle^2\tau_\text{r}^2\left(1-{\mathrm e}^{-t/\tau_\text{r}}\right)
  \nonumber \\
  &&\qquad
  - 6\frac{k_\text{B}T}{M_\text{m}}\tau_\text{v}^2\left(1-{\mathrm e}^{-t/\tau_\text{v}}\right)
\end{eqnarray}
Here, $\tau_\text{r}$ and $\tau_\text{v}$ are the reorientation and
velocity relaxation times, and $M_\text{m}$ is the motor mass.
The effective dimer diffusion coefficient is
$D_\text{m}=D_0+\frac{1}{3}\langle V_z\rangle^2\tau_\text{r}$, where
$D_0 = (k_\text{B}T/M_\text{m})\,\tau_\text{v}$.
In the ballistic regime, $t\ll\tau_\text{v}$, $\Delta L^2(t) \approx
(3k_\text{B}T/M_\text{m} + \langle V_z\rangle^2)\,t^2$.

The majority of research has been carried out on chemically self-propelled
motors with linear dimensions of microns or hundreds of nanometers,
similar to those of many swimming microorganisms.
Typical motor velocities are in the range of tens of micrometers per
second but could be even higher.
Given the motor speeds and sizes, and the kinematic viscosity of water,
this places these motors in the low Reynolds number regime.

If one scales down by 2 to 3 orders of magnitude to the regime where
motor linear dimensions are a few nanometers, the effects of fluctuations
are a dominant factor, and it is interesting to investigate the dynamical
properties of these tiny motors.
They now have sizes comparable to those of many protein motors and machines
in the cell.
Experimental observations of artificial self-propelled motors at the
molecular scale, Janus particles of 30 nm size~\cite{lee:14}
down to organometallic motors of 5 {\AA} size~\cite{pavlick:13},
show enhanced diffusion even at this small length scale.

\begin{figure}[tb]
  \centering
  \includegraphics{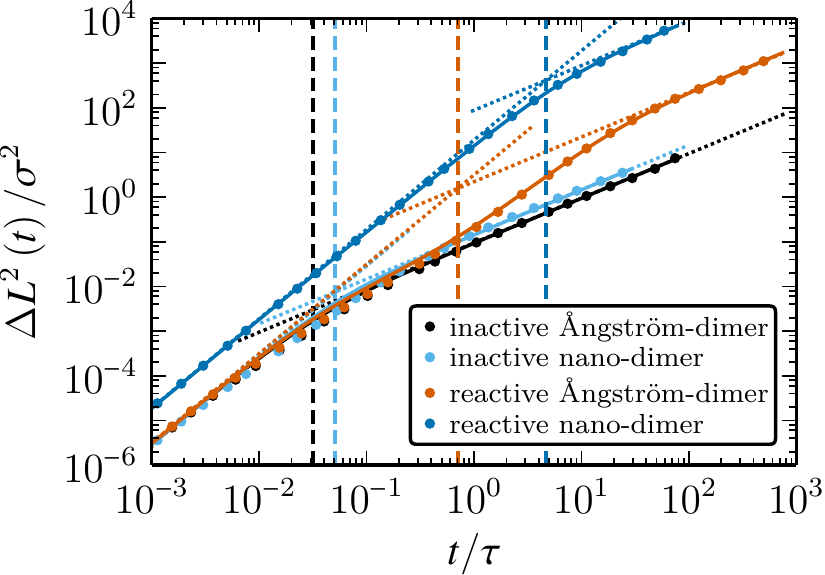}
  \caption{Mean-square displacements of inactive and reactive
  {\AA}ngstr{\"o}m-scale and nanoscale sphere dimers.
  The points show simulation data, and the solid lines show theoretical
  predictions.
  The vertical dashed lines indicate crossover times between the
  short-time ballistic and the long-time diffusive regimes.
  The data is shown in dimensionless time and length units of $\tau =
  \sigma\sqrt{M_\text{m}/k_\text{B}T}$ and $\sigma = \sqrt[3]{d_\text{C}^3 +
  d_\text{N}^3}$, respectively, where $d_\text{C}$ and $d_\text{N}$
  are the diameters of the C and N spheres.}
  \label{fig:dimer_msd}
\end{figure}
{\AA}ngstr{\"o}m-size chemically powered motors~\cite{colberg:14} have
been studied theoretically using molecular dynamics.
These tiny motors employ the same mechanism of propulsion through
diffusiophoresis as that of larger nanoscale and micron-size motors; in this
regime, however, whose motion is subject to very strong fluctuations and
solvent structural effects given the comparable sizes of motor and
solvent molecules.
The mean-square displacements of an {\AA}ngstr{\"o}m-size sphere dimer
and a nanoscale sphere dimer are shown in Figure~\ref{fig:dimer_msd},
where time and length have been rescaled to account for their different
sizes and thermal velocity.
For both length scales, the simulation data is in good agreement with
the theoretical prediction in eq~\eqref{eq:MSD-th}.
The rescaled MSDs of the inactive dimers, their motion is subject only
to thermal fluctuations, is approximately equal and thus independent
of the length scale.
The rescaled MSD of the reactive dimers, however, reveals significant
differences depending on the length scale.
In the ballistic regime, $t\ll\tau_\text{v}$, the MSD of the reactive nanodimer
is significantly larger than that of the inactive nanodimer, since
the thermal velocity $(k_\text{B}T/M_\text{m})^{1/2}$ is negligible
compared to the average propulsion velocity $\langle V_z\rangle$,
and $\Delta L^2(t)\approx\langle V_z\rangle ^2 t^2$.
The MSD of the reactive {\AA}ngstr{\"o}m-dimer, however, is almost equal
to that of the inactive {\AA}ngstr{\"o}m-dimer, since now the propulsion
velocity is negligible compared to the thermal velocity, and
$\Delta L^2(t) \approx (3k_\text{B}T/M_\text{m})t^2$.
In the diffusive regime, $t\gg\tau_\text{r}$, both the reactive
{\AA}ngstr{\"o}m-dimer and the reactive nanodimer exhibit enhanced
diffusion, where the enhancement is smaller (but still significant)
for the {\AA}ngstr{\"o}m-dimer, as expected due to the stronger thermal
fluctuations.
The crossover regime from the ballistic to the diffusive regime is short
for the nanodimer; for the {\AA}ngstr{\"o}m-dimer, however, the crossover
regime spans 3 orders of magnitude in (rescaled) time.
Given $\langle V_z\rangle$ and $\tau_\text{r}$, the average linear distance
traveled by a motor can be estimated as $\langle V_z\rangle\tau_\text{r}$.
The reactive nanodimer travels an average distance of 11.6 times its
effective diameter, $\sigma$; the {\AA}ngstr{\"o}m-dimer travels 3.0
times its effective diameter.
Thus, chemically powered motors can operate on very small length scales
and yield substantially enhanced diffusion coefficients, consistent
with recent experiments on active enzymes~\cite{sen-enz:10,sengupta:13}
and small Janus-like particles~\cite{lee:14}.
These observations suggest possible applications using very small
synthetic motors.

\section{Conclusions}
Investigations of small chemically powered motors present challenges
for experiment and theory.
They may also provide a diverse and transformative range of tools for
new applications.
Experimental challenges center around the design and construction of
micron and nanoscale motors with specific geometries, fueled by various
chemicals, operating by propulsion and control mechanisms selected for
specific purposes.
Since motors that might be used for some tasks may be very small,
continuum descriptions, while often applicable on surprisingly small
length and time scales, may nevertheless break down, and this
necessitates the use of microscopic or mesoscopic theories of the
dynamics.
Self-propelled motors function under nonequilibrium conditions, and their
full statistical mechanical description must account for the fluxes that
drive the system out of equilibrium.
In far-from-equilibrium regimes, systems display features, such as
bistability, oscillations, and self-propulsion, that are distinct from
those of equilibrium systems.
The statistical mechanics of driven nonequilibrium systems is a topic
of current research, and complete studies of chemically powered motors within
this context have not yet been carried out.

The potential uses of nanomotors have been discussed often in articles and
reviews, and proof-of-principle experiments have shown that operations,
such as cargo transport and motor-aided microfluidic flows, may soon
lead to viable applications.
Other applications, particularly those that involve biological systems,
will require the development of motors that use biocompatible fuels
and motor components.
In most cases, a single nanomotor is insufficient to complete a task.
A full understanding of the factors that lead to the collective behavior
of motors, the spatiotemporal structures that develop, and methods needed
to control ensembles of interacting motors must be achieved before many
applications can be carried out.
When such issues concerning motor design and control are completely
understood, it is possible that synthetic motors and active transport will
play as significant a role as molecular motors and machines currently play
in living systems.

\subsection*{Author Information}

\subsubsection*{Biographies}

\paragraph{Peter H. Colberg} received his Diplom in physics from the
Ludwig-Maximilians-Universit{\"a}t M{\"u}nchen.
He is currently a Ph.D. candidate at the University of Toronto, where he
studies chemically powered motors using microscopic and mesoscopic models.
He composes large-scale molecular simulation programs for highly parallel
processors such as GPUs.

\paragraph{Shang Yik Reigh} received his Ph.D. from Seoul National University.
He has postdoctoral research experience from the Forschungszentrum J{\"u}lich
and the University of Toronto.
His research interests are in the theory and simulation of diffusion-controlled
reactions, polymer melts, bacterial locomotion, and self-propelled nanomotors.

\paragraph{Bryan Robertson} received his B.Sc. in Chemistry from the
University of Western Ontario in 2013.
He is currently a Ph.D. student at the University of Toronto, studying the
behavior of nanomotors in complex media.

\paragraph{Raymond Kapral} received his Ph.D. from Princeton University and
pursued postdoctoral studies at the Massachusetts Institute of Technology.
He is currently Professor of Chemistry at the University of Toronto.
His research interests lie in the areas of nonequilibrium statistical
mechanics and quantum dynamics.

\subsection*{Acknowledgments}
This work was supported in part by a grant from the Natural Sciences and
Engineering Research Council of Canada and Compute Canada.

\bibliography{acr_nanomotor}

\end{document}